# Effect of impurities on pentacene island nucleation

*By B. R. Conrad, Elba Gomar-Nadal, W. G. Cullen, A. Pimpinelli[†], T. L. Einstein, E. D. Williams*

Physics Department and Materials Research Science and Engineering Center,
University of Maryland
College Park, MD 20742-4111, USA

[†] Also at LASMEA, UMR 6602 CNRS/Université Blaise Pascal -- Clermont 2,F-63177 Aubière, France

Corresponding author: edw@umd.edu

Pentacenequinone (PnQ) impurities have been introduced into a pentacene source material in a controlled manner to quantify the relative effects of the impurity content on grain boundary structure and thin film nucleation. Atomic force microscopy (AFM) has been employed to directly characterize films grown using 0.0-7.5% PnQ by weight in the source material. Analysis of the distribution of capture zones areas of submonolayer islands as a function of impurity content shows that for large PnQ content the critical nucleus size for forming a Pn island is smaller than for low PnQ content. This result indicates a favorable energy for formation of Pn-PnQ complexes, which in turn suggests that the primary effect of PnQ on Pn mobility may arise from homogeneous distribution of PnQ defects.

The study of organic materials, particularly the various roles of morphology and impurity doping, remains an active subject for device physics, materials design, and applied statistical mechanics.[1-3] Studies of the most promising organic electronic semiconductor, pentacene (Pn), have shown that its transport properties are sensitively dependent on crystalline quality[4,5] and thin film preparation: for the work here, observations that low concentrations of impurities significantly affect film nucleation and growth, electronic transport, and electronic signal noise are of particular interest [6-9]. Extensive studies of the initial stages of pentacene film growth [3, 10-15] have shown that it follows the classical picture of nucleation, island growth, aggregation and coalescence that was developed for the growth of inorganic films.[16-20] In later stages of growth, the two-dimensional domains formed from island coalescence serve as the basis for three-dimensional growth due to an Ehrlich-Schwoebel energy barrier that slows diffusion from higher to lower layers of the film [5, 7, 21]. Scaling analysis has proven powerful for evaluating island nucleation and grain boundary formation in such growth systems.[17, 19, 20, 22] In particular, recent investigations using the Wigner surmise, which relates growth processes to universal aspects of fluctuations, have yielded significant improvements in physical understanding. [17, 20] We have measured changes in the capture zone distributions for Pn films grown in the presence of low impurity concentrations, and use the Wigner analysis to demonstrate that the underlying cause is an impurity-induced decrease in the number of molecules required to form a critical nucleus.

The experiments were performed by introducing controlled amounts of the chemical impurity – 6,13-pentacenequinone (PnQ) – into pentacene (Pn). The solid mixtures were prepared by mechanically mixing under a dry nitrogen atmosphere. A series of films were prepared on highly doped Si (100) wafers with 300nm thermally grown oxide pre-cleaned using standard procedures[7] based on many years of experience in preparing atomically cleaned Si samples[23, 24]. The source materials were increased to the deposition temperature (195ºC) over a 15 minute interval. Deposition was performed at 0.09 Å/s at $10^{-7}$ Torr pressure, with the substrate at room temperature. The compositions tested covered a range of added PnQ from weight percentage +0.0 to 7.5%, equivalent to added PnQ number fractions ranging from 0.000 to 0.068. The added impurity supplements the natural impurity level of commercial Pn, which is approximately 0.7% by weight or a number fraction of 0.006, as determined previously.[5, 7] To prepare materials with lower impurity content, source material was heated to a temperature slightly lower than its sublimation temperature for at least one hour *prior* to the thin film deposition. Previous measurements have shown that this treatment reduces the absolute source PnQ number fraction to less than 0.001 [7], and yields sample mobilities as high as those obtained with Pn purified using gradient-sublimed material. The source concentration values used to quantify our results are the added number fraction plus the natural impurity level. This represents a readily reproducible quantity, but will not represent the absolute concentration in the thin film, due to the larger sublimation rate of PnQ than Pn at any given source temperature. Two film thicknesses were grown, submonolayer and 50 nm thick, with two different growths for each thickness. The film morphology was characterized using tapping-mode AFM. The islands in the submonolayer thin-films are quantified in image processing by setting a height threshold to account for substrate height distribution. A limit is also placed on the minimum areal island size to account for image noise. Voronoi polygons (Wigner-Seitz cells) are then calculated from the island nucleation data. For thick films the grain sizes were found by using automated routines to outline the irregularly shaped grains and measure their areas.

AFM images of a subset of the prepared submonolayer films as a function of the source number fraction of the PnQ impurity of the source material are shown in Fig 1. As the impurity content of the source material is increased, the films display PnQ phase-separation growth, characterized by the appearance of tall islands (that appear as white areas in the AFM images). Previous studies have shown that these tall islands are crystalline PnQ.[2, 6, 7, 25] Sample AFM images of thick films, displayed in Fig. 2, show that increasing concentration of PnQ during growth causes an abrupt change for PnQ number fractions larger than 0.008 in the ultimate grain sizes and local structure of the bulk film. There is substantial local variation in the shape and sizes of the grains across a sample. The sizes of the pentacene grains were thus measured as averages over three or more images for each deposition and, as Fig. 4a and Table I show, both the thick-film average pentacene grain size and the submonolayer average island size decrease abruptly when the impurity concentration reaches a number fraction ~0.008 PnQ. For grain size determination, the tall PnQ growths were excluded. This decrease is concurrent with large variations in individual grain size as well as the PnQ phase separation shown in Fig. 2. Substantially decreased electrical transport performance begins well before the observable morphological changes and the region where grain size is decreasing coincides with a further factor-of-four decrease in the material's mobility.[7] Several mechanisms by which impurities could be incorporated into the pentacene film and limit transport have been proposed in the literature, including changes in chemical bonds, disruption of the crystalline structure within a grain, and impurity accumulation at the grain boundaries.[7, 15, 26-28] In the following, careful analysis of the growth mode changes due to the PnQ is used to help understand where the PnQ resides in the thin films, and thus clarify the mechanism by which PnQ reduces the mobility of the Pn.

Traditionally, nucleation studies have characterized the evolving submonolayer growth in terms of the island-size distribution (ISD), which under general circumstances has a coverage-insensitive form dependent only on the ratio of the island size to its mean. Another metric monitors the distribution of capture zones (CZ).[2, 17, 29] These CZs are essentially the proximity (Wigner-Seitz) cells of the islands: the CZ is the number of sites (times the area associated with each) that are closer to the enclosed island than to any other island. Thus, CZs are essentially the areas of Voronoi polygons that are created from the island nucleation points. The capture zone distribution (CZD) can be similar to the ISD but also may differ even qualitatively, particularly for slow deposition. It was recognized [30-32] over a decade ago that analyzing the CZ distribution (CZD) can be more fruitful than the ISD, which also tends to be more sensitive to deposition rate. Application of the CZ analysis is illustrated in Fig. 3, which shows a 10 $\mu$m $\times$ 10 $\mu$m AFM image of a 0.3 monolayer commercial pentacene deposition with the centers of the islands and calculated Voronoi polygons indicated by black dots and lines respectively.

Various formal expressions have been used to characterize the CZD, the simplest of which is a gamma distribution.[20, 31-34] Recently some of us have shown that the generalized Wigner distribution (GWD) accounts for experimental or Monte Carlo data comparably to, if not better than, the gamma distribution and reveals, as described below, fundamentals of the nucleation process.[17] The GWD has the explicit form

$$P_\beta(s) = a_\beta s^\beta \exp(-b_\beta s^2) \qquad (1)$$

where $s$ is the CZ area normalized by the mean CZ area. The exponent β is the only free parameter, and its value is directly related to the critical nucleus size (see below), while $a_\beta$ and $b_\beta$ are (β-dependent) constants determined by normalization and unit mean, respectively.[35] A representative example of the fit of a CZD by the GWD is shown in Fig. 4b. The inset of Fig 4b and Table I give the exponent β as a function of the level of source impurity PnQ content. We find β = 4.97 ± 0.26 for the CZD for number fractions between 0.008 and 0.052, indicated by the solid line. The CZDs at lower concentrations of PnQ have exponent β = 6.65 ± 0.26. The width of the distribution[36] $\sigma = \sqrt{(\beta+1)/(2b_\beta) - 1}$, follows the opposite trend with σ = 0.260 ± 0.004 for $N > 0.008$ and σ = 0.295 ± 0.007 for $N \leq 0.008$. In contrast, if we fit the ISDs with the GWD, the average value of β (and the corresponding width of the distribution) is insensitive to added impurity content, as summarized in Table 1.

In two dimensions the characteristic exponent β = $i$ + 1, where $i$ is the critical nucleus size (i.e., $i$ + 1 is the number of adspecie particles in the smallest stable island). The values of the exponent β, therefore indicate a change in the critical nucleus size, from $i \sim 6$ when the impurity content is small, to $i \sim 4$, when the impurity content is large.

In a study of the ISDs of pure Pn films of fractional coverages 0.18 and 0.42, Ruiz et al. found that the ISD of the two overlayer densities collapsed onto a single scaling curve in normalized island size. Using Amar and Family's semiempirical expression,[19] they showed that the critical nucleus was decidedly larger than a point island or a dimer. While their distribution was notably noisier than that in our Fig. 4b, their least-squares fit gave $i$ = 3, to be compared with the value 3.8 ± 0.2 reported in Table 1. The comparison is as expected, because the values of $i$ predicted by ISDs are consistently lower than those obtained from CZDs. Both Monte Carlo simulations and experiments, especially those on Pn[11], have shown that the CZD is more robust than the ISD. Fig. 3 of Ref. 11 provides a convincing illustration: The CZD is insensitive to the deposition rate (relative to surface diffusion); for rapid deposition of Pn the ISD is comparable to the CZD, but for slower deposition the ISD rises more rapidly to a maximum at smaller normalized size[11]. When fit with the GWD, Eq. 1, such behavior corresponds to a smaller value of *i*.

A plausible explanation for the dependence of critical nucleus size on impurity content would be the existence of preferential interaction between PnQ and Pn molecules, allowing small clusters to form with greater stability. This mechanism would be likely to result in the inclusion of a low density of PnQ within the grains of Pn, providing an explanation for the strong decrease in mobility observed at very low number density of PnQ,[7] even though the grain boundary distribution has not changed observably. While models for Pn transport have focused on grain boundary defects and impurities,[28, 37, 38] impurities within the grain could also cause gap states and thus charge traps,[26, 27] or generate hole scattering similar to that at grain boundaries.[28] Some experimental evidence suggests that trap states are homogeneously distributed in Pn thin films,[39, 40] although no attempt has yet been made to correlate these observations with impurity content.

The stabilization of the critical nucleus size above a PnQ number density of ~0.008 is consistent with the coexistence of a disordered and crystalline phase of PnQ, with an equilibrium density in the disordered phase of ~0.008, as previously reported[7]. The complete absorption of excess PnQ into the crystalline phase would occur when the diffusion length of a PnQ molecule is larger than the separation of PnQ islands. This suggests that sequential, rather than mixed, deposition of Pn and PnQ may directly reveal density-dependent nucleation of islands. Figs. 5a and 5b show a $20\,\mu m \times 20\,\mu m$ AFM image and a $10\,\mu m \times 10\,\mu m$ enlargement, respectively, of the same image of a film grown by depositing 3.6 Å of PnQ followed by 3.2 Å of Pn. Areas near large PnQ crystallites contain a lower density of larger Pn islands than locations far away from the large PnQ crystallites. This suggests that the nucleation of pure PnQ crystallites lowered the local density of PnQ to the equilibrium density, while areas where PnQ crystallites did not nucleate were left with a local excess of molecular PnQ, which enhanced the nucleation of Pn islands. In addition, it is possible that a preferential interaction between Pn and PnQ would cause Pn near PnQ islands to coat the PnQ crystallites, thus reducing the Pn density available to form islands. It is not possible to differentiate between these two possibilities with the stated series of experiments nor are the effects mutually exclusive.

In summary, the morphology of co-deposited submonolayer films has been analyzed in terms of the capture zone and island size distributions. The distributions are well described using the generalized Wigner distribution. The greater sensitivity of the capture zone distribution to growth processes reveals that the critical nucleus size for Pn island formation decreases for source-PnQ number fractions larger than ~0.006, from $i \sim 6$ (fitting parameter = $5.65 \pm 0.25$) at lower PnQ density to $i \sim 4$ (fitting parameter = $3.97 \pm 0.26$), suggesting PnQ enhances the formation of molecular complexes that can serve as nucleation sites. Increased impurity doping primarily results in continued phase-separation with diffusion driven differences in nucleation density.

Work at the University of Maryland was supported by NSF-MRSEC Grant DMR 05-20471 and use of MRSEC SEFs, by the LPS and by NIST under contract

#70NANB6H6138, with research infrastructure supported by the UMD-CNAM and the UMD NanoCenter. Visits to Maryland by A.P. were supported by a CNRS Travel Grant, and T.L.E. was partially supported by DOE CMSN grant DEFG0205ER46227.

Figures

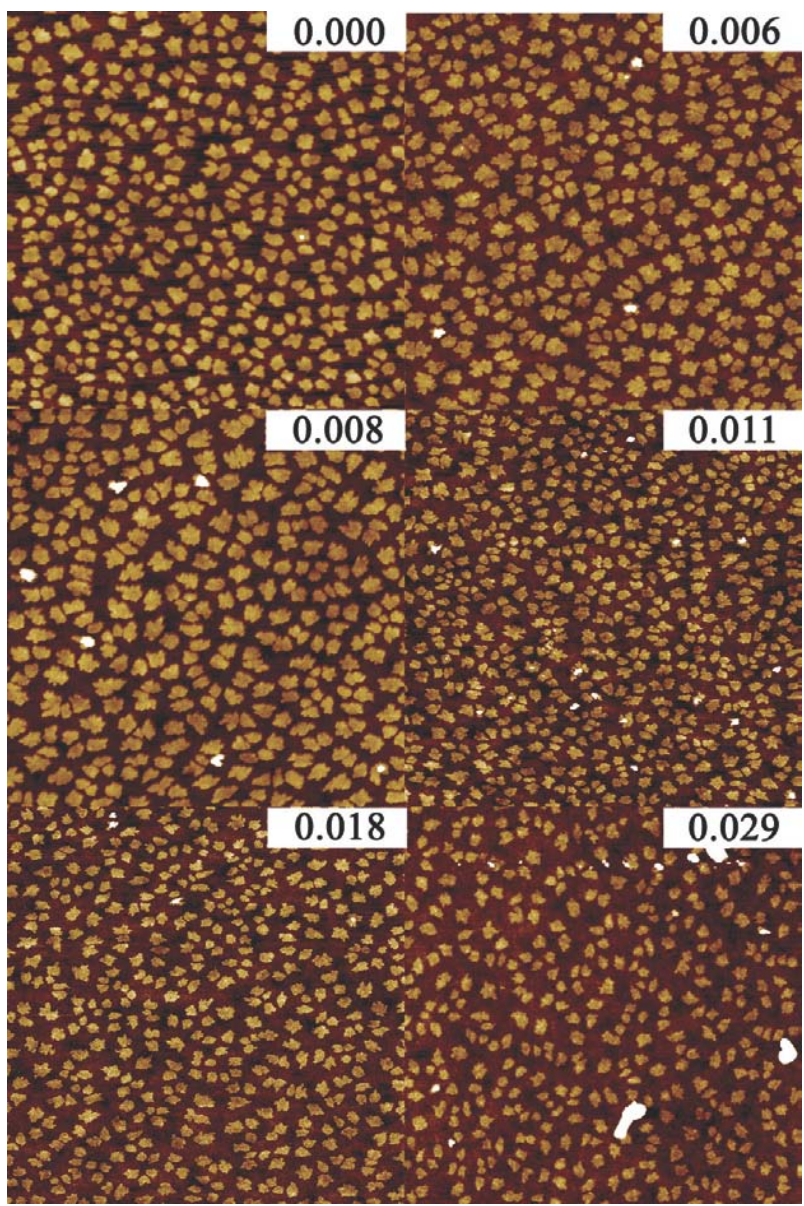

**Figure 1**: (Color online) AFM images ($10\ \mu\text{m} \times 10\ \mu\text{m}$) of 0.3 ML Pn/PnQ films on SiO$_2$ with varying source PnQ number fractions.

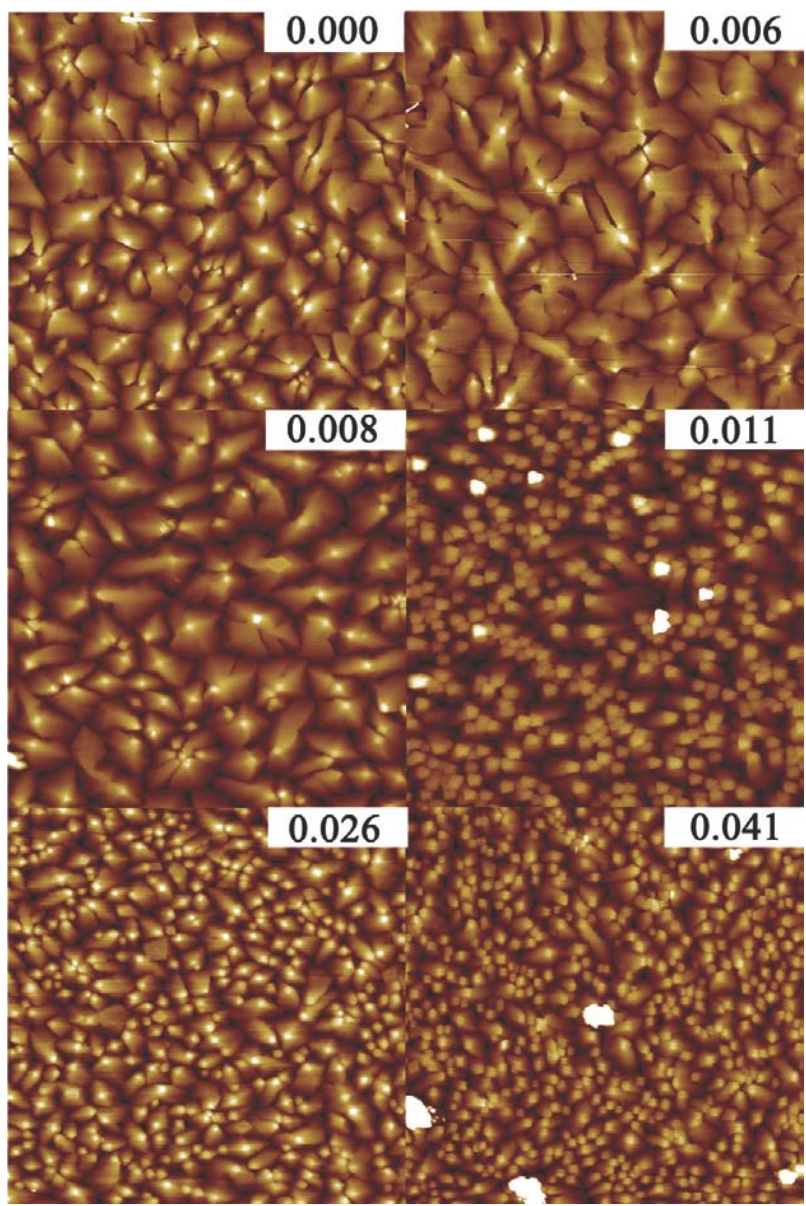

**Figure 2**: (Color online) AFM images ($10\,\mu\text{m} \times 10\,\mu\text{m}$) of 50 nm Pn/PnQ films on $SiO_2$ with source materials containing varying PnQ number fractions. Note the varying grain size and morphology.

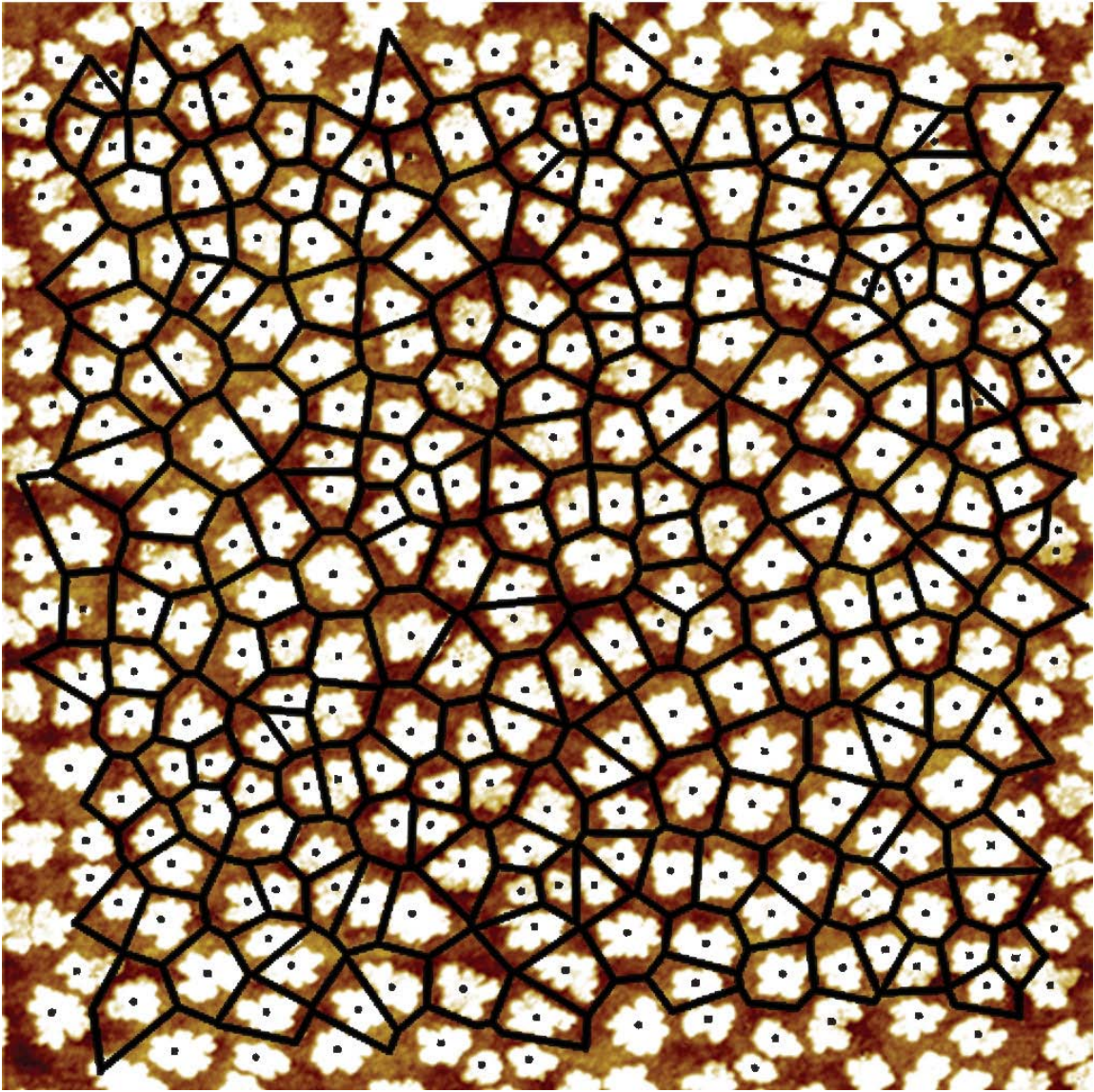

**Figure 3**: (Color online) An example (10 $\mu$m $\times$ 10 $\mu$m) AFM image of commercial Pn. The island centers and Voronoi polygons are indicated by black dots and lines, respectively.

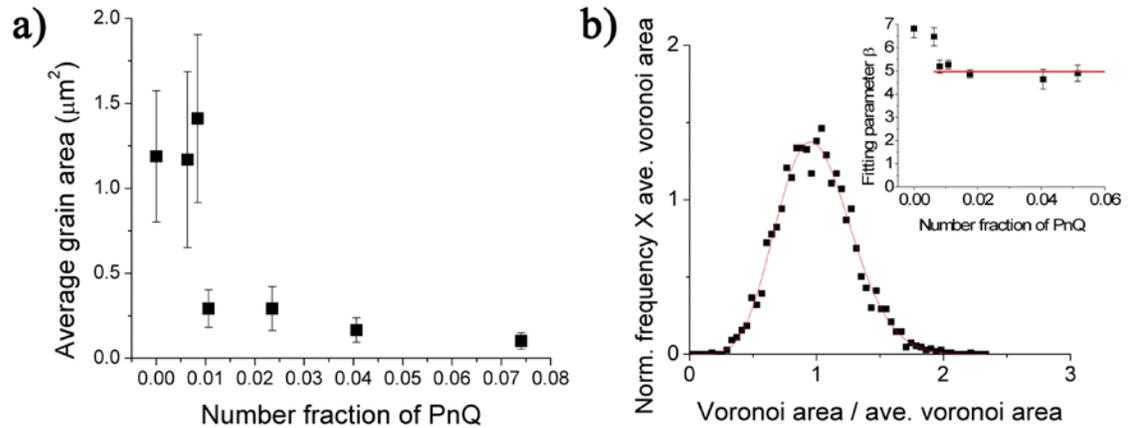

Figure 4: (Color online) **a)** Average thick film grain area as a function of the number fraction of PnQ content of the source material. **b)** An example of the normalized CZ area histogram. The solid line is the generalized Wigner surmise distribution fit with β= 5.27 ± 0.19. The inset is the capture zone fitting parameter ρ as a function of the number fraction of PnQ content of the source material.

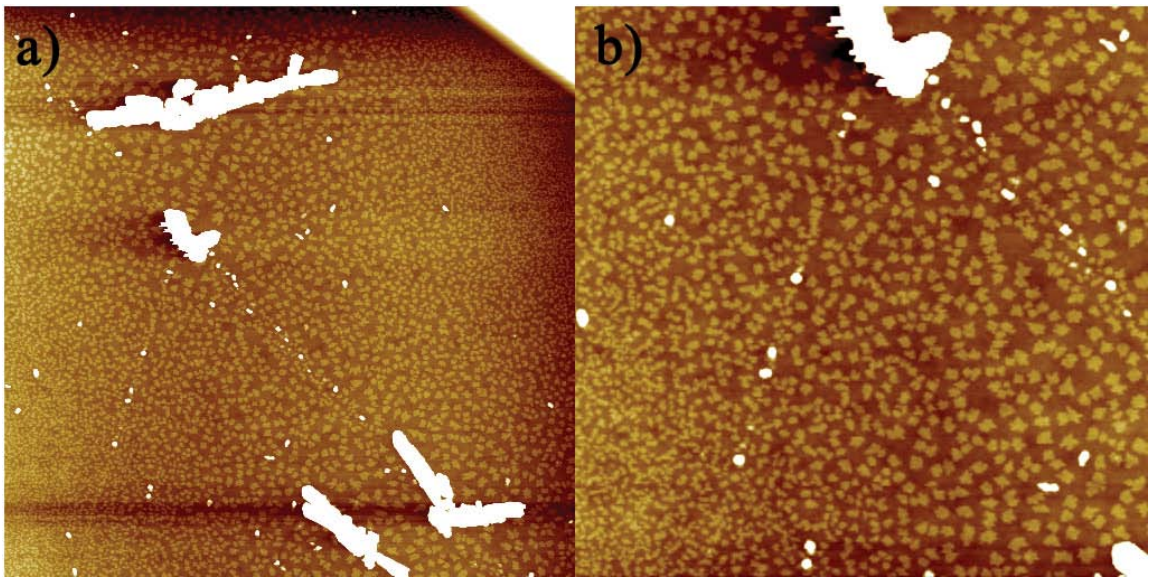

**Figure 5**: (Color online)  a) A ($20\,\mu m \times 20\,\mu m$) AFM of image of a sample with 3.6Å PnQ deposition followed by a 3.2Å Pn deposition illustrating larger islands near the large topological features. b) A ($10\,\mu m \times 10\,\mu m$) zoom-in of Fig. 5a.

**Table 1:** The average submonolayer island size in $\mu m^2$ (AIS), mean capture zone area in $\mu m^2$ (MCZ), capture zone distribution Wigner exponent β (CZD-β), the island size distribution Wigner exponent β (ISD-β), and average thick film grain size in $\mu m^2$ (AGS) as a function of the number fraction of PnQ.

| PnQ Number Fraction | Average Island Area (AIS) | Capture Zone Area (MCZ) | Capture Zone Exponent - β | Island Area Exponent - β | Average Grain Area (AGS) |
|---|---|---|---|---|---|
| 0.000 | 0.12±0.06 | 0.36±0.11 | 6.8±0.4 | 3.8±0.2 | 1.2±0.4 |
| 0.006 | 0.12±0.04 | 0.30±0.08 | 6.5±0.4 | 4.4±0.3 | 1.2±0.5 |
| 0.008 | 0.12±0.04 | 0.31±0.09 | 5.2±0.3 | 3.6±0.2 | 1.4±0.5 |
| 0.011 | 0.036±0.016 | 0.13±0.04 | 5.3±0.2 | 3.9±0.2 | 0.29±0.11 |
| 0.018 | 0.065±0.022 | 0.22±0.06 | 4.9±0.2 | 3.9±0.2 | 0.29±0.12 |
| 0.041 | 0.068±0.026 | 0.26±0.08 | 4.6±0.4 | 4.1±0.4 | 0.17±0.07 |
| 0.052 | 0.083±0.029 | 0.30±0.09 | 4.9±0.4 | 4.2±0.4 | 0.10±0.05 |